\documentclass[twocolumn,showpacs,preprintnumbers,amsmath,amssymb,prb]{revtex4}

\usepackage[dvips]{graphicx}
\usepackage{dcolumn}
\usepackage{bm}
\usepackage{hyperref}
\hypersetup{colorlinks=blue}

\begin{document}

\preprint{APS/123-QED}

\title{Carrier Localization, Metal-Insulator Transitions and Stripe Formation in Inhomogeneous Hole-Doped Cuprates}

\author{S. Dzhumanov}
\email{dzhumanov@rambler.ru} \affiliation{%
Institute of Nuclear Physics, 100214, Ulughbek, Tashkent,
Uzbekistan}

\author{O.K. Ganiev}
\affiliation{%
Institute of Nuclear Physics, 100214, Ulughbek,
Tashkent, Uzbekistan}

\author{Z.S. Khudayberdiev}
\affiliation{%
Institute of Nuclear Physics, 100214, Ulughbek,
Tashkent, Uzbekistan}

\date{\today}

\begin{abstract}
We propose a unified approach for describing the carrier
localization, metal-insulator transitions (MITs) and stripe
formation in inhomogeneous hole-doped cuprates. The ground-state
energy of a carrier interacting with a defect and with lattice
vibrations is calculated variationally within the continuum model
and adiabatic approximation. At low doping levels, hole carriers in
$\rm{La}$-based systems with large-radius dopants are localized near
the dopants with the formation of hydrogen-like impurity centers. As
the doping increases, the carriers are liberated from the
hydrogen-like impurity centers and are self-trapped in a defect-free
deformable lattice with the formation of intrinsic large polarons.
In $\rm{La}$-based cuprates with small-radius dopants, hole carriers
are self-trapped near the dopants with the formation of
non-hydrogen-like impurity centers or extrinsic large polarons. The
charge ordering and formation of different superlattices and energy
bands of dopants and intrinsic large polarons at their inhomogeneous
spatial distribution trigger the MITs in cuprates. We analyse the
validity of the criteria for the Mott and Anderson MITs in
$\rm{La}$-based cuprates with hydrogen-like impurity centers and
show that such MITs in these systems are unlikely possible. Using
the uncertainty principle we derive the quantitative criteria for
the new MITs caused by strong carrier-dopant-phonon and
carrier-phonon interactions. We show that the MITs in
$\rm{La}$-based cuprates with large-radius dopants are driven by
intrinsic self-trapping and charge ordering, while the anisotropy of
dielectric constants fosters the MITs accompanied by the formation
of static stripes (in carrier-poor domains) and dynamic ones (in
carrier-rich domains) in the range of doping $x\simeq$0.04-0.125.
The so-called 1/8 anomaly is especially pronounced in these systems.
The small-radius dopants and the anisotropy of dielectric constants
favor carrier localization, MITs and stripe formation in a wide
range of doping (including also "magic" doping $x$=1/8) in other
$\rm{La}$-based cuprates, where $x$=1/8 is no longer "magic" doping.
Finally, the MIT and 1/8 anomaly in cuprates caused by the
commensurate ordering of planar large bipolarons and their
condensation into a liquid are also discussed. The obtained results
are in quantitative agrement with the experimental data on polaron
formation, MITs and stripe formation in $\rm{La}$-based cuprates.
\end{abstract}

\maketitle

\section{Introduction}\label{sec:level1}

The carrier localization, metal-insulator transitions (MITs) and
stripe formation in hole-doped cuprates are the most challenging
problems in condensed matter physics. These phenomena have attracted
much attention because they are closely related to high-$T_c$
superconductivity in cuprates
\cite{1,2,3,4,5,6,7,8,9,10,11,12,13,14,15}. The mechanisms for
carrier localization and MITs in the cuprates and related materials
have been under discussion since the discovery of high-$T_c$
superconductivity \cite{1,2,3,4,5,6,7,8,9,10,12,15,17,18,19}.
Another important problem of the physics of high-$T_c$ cuprates is
the role of the electronic inhomogeneity and charge ordering in the
stripe formation \cite{11,14,20,21}, which is intimately related to
carrier localization and MITs in these materials. For cuprates
within the insulating antiferromagnetic (AF) phase, the electron
correlation is likely to dominate the carrier localization; when a
MIT is approached, it is not obvious which process will dominate
\cite{12,22}. Actually, the MITs in the cuprates might be different
from the Mott and Anderson transitions. In these systems, the
phenomena of carrier localization seem to be much more complicated
by electron-defect-lattice and/or electron-lattice interaction
effects which become very pronounced close to the MIT \cite{9,22}.
Generally, the electron-defect-lattice and/or electron-lattice
interactions and the electronic inhomogeneity are known to foster
carrier localization in solids. In particular, the strong
electron-defect-phonon and electron-phonon interactions foster
carrier localization by causing extrinsic and intrinsic
self-trapping. At the same time the gross electronic inhomogeneities
might play an important role in carrier localization, MITs and
formation of insulating and metallic stripes in the cuprates. There
is now increasing experimental evidence for the existence of
extrinsic polarons (i.e., charge carriers self-trapped near the
dopants or impurities), intrinsic polarons (or charge carriers
self-trapped in a defect-free deformable lattice) and electronic
inhomogeneities in these materials \cite{9,20,21,23,24}. In
addition, various experiments on hole-doped $La$-based cuprates show
that the MITs and stripe formation occur in underdoped (at doping
levels $x=0.05-0.125$) \cite{25,26}, optimally doped ($x=0.15-0.16$)
\cite{8}, and overdoped ($x=0.16-0.25$) \cite{16,21,27} regimes.
However, the role of electronic inhomogeneities and strong
carrier-defect-phonon and carrier-phonon interactions, which is
known to be very important in the vicinity of the MITs
\cite{20,22,28}, has not been fully explored in the underdoped
regime. While the effects of ionic (or dopant) size and anisotropy
of dielectric constants, which may play a key role in carrier
localization and MITs, have not been studied theoretically in
inhomogeneous $\rm{La}$-based cuprates in a wide hole concentration
range from $x\sim0.02$ (lightly doped regime) to $x\sim0.25$
(heavily overdoped regime). In this work we examine such
inhomogeneous systems in which the electronic charge inhomogeneities
in conjunction with the strong carrier-defect-phonon and/or
carrier-phonon interactions induce the carrier localization and MITs
accompanied by the stripe formation in a wide range of doping. In so
doing, we study the MITs and stripe formation in high-$T_c$ cuprates
driven by the inhomogeneous local doping, charge ordering and
self-trapping of hole-carriers near defects (impurities) or in a
defect-free deformable lattice.

This paper is organized as follows. First, in Sec. II, we
investigate the extrinsic (or defect-assisted) and intrinsic
self-trapping of carriers, and analyze the possibility of formation
of extrinsic (hydrogen-like and non-hydrogen-like) and intrinsic
self-trapped states in hole-doped $\rm{La}$-based cuprates. Then, in
Sec. III, we consider the possible forms of charge ordering and the
formation of different superlattices and energy bands of dopants
(i.e., acceptors) and intrinsic large polarons at their
inhomogeneous spatial distribution. In Sec. IV, we examine the
validity of the criteria for the Mott and Anderson transitions in
$\rm{La}$-based cuprates with hydrogenic impurity centers. We argue
that such transitions in the cuprates are unlikely possible. We give
our view of the new MITs in $\rm{La}$-based cuprates with extrinsic
large polarons (non-hydrogen-like impurity centers) or intrinsic
large polarons and derive the quantitative criteria for the MITs by
using the pertinent uncertainty relation. We show that in these
systems the extrinsic and/or intrinsic self-trapping of hole
carriers and the distinctly different charge ordering result in the
new MITs which occur at different doping levels. Further, we find
that, the ionic radius of dopants and the anisotropy of dielectric
constants strongly influence on carrier localization, MITs, stripe
formation and so-called 1/8 anomaly in $\rm{La}$-based cuprates.
These predictions are consistent with distinctive features of MITs,
stripe formation and 1/8 anomaly observed in the cuprates. Finally,
we briefly discuss the other type of MIT and the 1/8 anomaly in
hole-doped cuprates caused by the commensurate ordering of planar
large bipolarons and their condensation into a liquid. The paper
concludes with Sec. V, in which the principle results are
summarized.
\section{Extrinsic and intrinsic self-trapping of hole carriers}\label{sec:level2}
The undoped layered cuprates are typical charge-transfer (CT)-type
Mott insulators. According to the band calculations and
spectroscopic data \cite{10,29}, the electronic structure of these
parent $\rm{CuO_2}$-based compounds is well described by the
three-band Hubbard model and the oxygen valence band lies within the
Mott-Hubbard gap. Upon hole doping, the oxygen valence band of the
cuprates is occupied by free holes. The hole carriers are assumed to
be within both a three-dimensional (3D) and two-dimensional (2D)
deformable medium, the last one being $\rm{CuO_2}$ layers
\cite{1,6}. The large static dielectric constant $\varepsilon_0$
$(>>10)$ both parallel and perpendicular to the $\rm{CuO_2}$ layers
\cite{1,9} indicate that lattice ions may be significantly displaced
in all three directions. In reality, the $\rm{CuO_2}$ based
materials may be approximated as a 3D deformable medium \cite{1}. As
it was pointed out in Ref. \cite{30}, the existence of the 3D
bismuth oxide high-$T_c$ superconductors, which belong to the same
class of materials as the copper oxides, casts an extra doubt on the
theories based mainly on the low dimensionality of the crystal
structure. There is also convincing experimental evidence that the
consideration of cuprates as 3D systems may appear to be more
appropriate (see Ref. \cite{12,31}). In these anisotropic 3D polar
materials, the hole carriers interacting both with lattice
vibrations (i.e., acoustic and optical phonons) and with lattice
defects (e.g., dopants or impurities), can easily be self-trapped
near the lattice defects. Such extrinsic self-trapping of hole
carriers leads to the formation of a moderately deep extrinsic
polaronic or a shallow hydrogenic state in the CT gap of the parent
cuprates. The theory of carrier self-trapping has been developed in
Refs. \cite{1,5,6,32,33} in the framework of the continuum approach.
As shown in Refs. \cite{33,34}, the continuum theory of carrier
self-trapping in a defect-free system can be easily extended to the
system consisting of a defect and a bound charge carrier. Such a
continuum approach is better suited for the quantitative analysis.
In the framework of the continuum model, we make an attempt to
analyze the possibility of the formation of hydrogen-like and
non-hydrogen-like impurity states as well as polaronic ones in the
CT gap of the cuprates.

The ground-state energy  of  a hole carrier in 3D polar crystals can
be calculated variationally in the continuum model and adiabatic
approximation, taking  into account the short- and long-range
carrier-defect-phonon interactions. The total energy of the
interacting carrier-defect-phonon system is given by the functional
\cite{34}
\begin{eqnarray}\label{1Eq}
E\{\psi(r)\}=-\frac{\hbar^2}{2m^\ast}\int\psi(r)\nabla^2\psi(r)d^3r-\nonumber\\
\frac{e^2}{2\tilde\varepsilon}\int\frac{\psi^2(r)\psi^2(r')}{|r-r'|}d^3rd^3r'-\nonumber\\
\frac{E_d^2}{2K}\int\psi^4(r)d^3r-\frac{Ze^2}{\varepsilon_0}\int\frac{\psi^2(r)}{r}d^3r+\nonumber\\
\left(V_0-\frac{E_dE_{dD}}{K}\right)\int\psi^2(r)\delta(r)d^3r ,
\end{eqnarray}
where $\psi(r)$ and $m^{\ast}$ are the wave function and effective
mass of a carrier, $\tilde\varepsilon=\varepsilon_{\infty}/
(1-\eta)$ is the effective dielectric constant,
$\eta=\varepsilon_{\infty}/\varepsilon_0$, $\varepsilon_\infty$ is
the high frequency dielectric constant, $K$ is an elastic constant,
$E_d$ and $E_{dD}$ are the deformation potentials of the carrier and
the defect, respectively, $V_0$ is the short-range defect potential,
$Z$ is the charge state of the defect.

Now, we may choose the trial wave function $\psi(r)$ in the Gaussian
form
\begin{equation}\label{2Eq}
\psi(r)=N\exp\left[-(\sigma r)^2\right]
\end{equation}
and make a variational calculation of Eq. (\ref{1Eq}) with respect
to the parameters contained in $\psi(r)$, where
$N=(2/\pi)^{3/4}\sigma^{3/2}$ is the normalization factor,
$\sigma=\sqrt{\pi}\beta/a_0$ is the variational parameter,
$\beta=a_{0}/a$ is the localization parameter, $a$ is the radius of
localization limited between $a_0$ (the lattice constant) and
$\infty$ (free carrier). Inserting Eq. (\ref{2Eq}) into Eq.
(\ref{1Eq}) and performing the $r$ integration in Eq. (\ref{1Eq}),
one can obtain the following functional
\begin{equation}\label{3Eq}
E(\beta)=B[\beta^2-g_1\beta^3-g_2\beta],
\end{equation}
where $B=3\pi\hbar^2/2m^{\ast}a^2_0$,
$g_{1}=(E^2_d/2Ka^3_0B)(1+b_s)$,
$b_{s}=2^{5/2}\left[E_{dD}/E_d-KV_0/E^2_d\right]$,
$g_{2}=(e^{2}/\varepsilon_{\infty}a_0B)(1-\eta+b_l\eta),b_l=2^{3/2}Z$.

When $3g_{1}g_{2}<1$ the functional (\ref{3Eq}) has a minimum at
$\beta_{min}=\left[1-\sqrt{1-3g_{1}g_{2}}\right]/3g_{1}$ which
corresponds to the formation of a large-radius impurity state or
extrinsic polaronic state in the CT gap of the doped cuprates. A
carrier interacting with lattice vibrations becomes a large polaron
with a polarization (or deformation) cloud extending over a wide
region. We can determine the ground-state energy of such an
intrinsic large polaron from Eq.(\ref{3Eq}) at $b_s$=0 and $Z=0$. At
$3g_{1}g_{2}<1$ the potential barrier height separating large- and
small-radius self-trapped states of a hole carrier is equal to
\cite{34,35}
\begin{eqnarray}\label{4Eq}
E_{a}=\frac{4B}{27g^{2}_{1}}\left[1-3g_{1}g_{2}\right]^{3/2}.
\end{eqnarray}

While the ground-state energy of a self-trapped carrier is given by
\cite{36}
\begin{eqnarray}\label{5Eq}
E(\beta_{min})=\frac{B}{27g^{2}_{1}}\left[2-9g_{1}g_{2}-2(1-3g_{1}g_{2})^{3/2}\right]
\end{eqnarray}

The signs of $E_d$ and $E_{dD}$ for holes and small-radius defects
always positive, while $E_{dD}$ for large-radius defects is negative
\cite{34}. In $\rm{La_{2-x}Sr_{x}CuO_{4}}$ (LSCO) the radius of
$\rm{Sr^{2+}}$ ions is larger than that of $\rm{La^{3+}}$ ions
\cite{37}, so that for $Sr^{2+}$ ion $Z=1$, $E_{dD}<0$ and
$b_{s}<0$. In this case, the short-range part of the impurity
potential in Eq. (\ref{1Eq}) is repulsive and the hole-lattice
interactions near such dopants are suppressed by this repulsive
defect potential. At a weak carrier-lattice interaction the
localized impurity state may have hydrogen-like character described
by a rigid lattice model \cite{38}. Therefore, in order to study the
Mott MIT in LSCO, we can consider the hydrogenic impurity centers
having the Bohr radius
$a_{H}=0.529\varepsilon_{0}(m_{e}/m^{*}){\AA}$ and the ionization
energy $E_I=e^{2}/2\varepsilon_{0}a_{H}$ in the Hubbard model, where
$m_e$ is the free electron mass. However, the substitution of
small-radius cations (e.g., $\rm{Ca^{2+}}$ and $\rm{Nd^{3+}}$ ions)
for $\rm{La^{3+}}$ ions in $\rm{La_{2}CuO_{4}}$ and for
$\rm{Sr^{2+}}$ ions in LSCO leads to the formation of the
non-hydrogenic impurity centers or extrinsic large polarons for
which the Hubbard model is already inapplicable. The distinctive
feature of the cuprates is their very large ratio of static to high
frequency dielectric constants \cite{1,39}. This situation is
favorable for carriers attracted to polarization well created by the
other ones or to Coulomb centers (dopants) to form 3D intrinsic or
extrinsic large bipolarons. Such bipolarons with small binding
energies can be formed in the lightly doped cuprates at $\eta<0.1$
\cite{36} and they become unstable in the underdoped regime.
\section{Possible types of charge ordering and Formation of different superlattices}\label{sec:level3}
The hole-doped cuprates are inhomogeneous or disordered systems,
where the dopants and charge carriers are distributed
inhomogeneously. Such dopant-driven and carrier-driven
inhomogeneities may produce regions with different doping levels.
The electronic inhomogeneity commonly exists in high-$T_{c}$
cuprates regardless of doping level and the underdoped cuprates
become more inhomogeneous than overdoped ones \cite{20}.
Inhomogeneous distributions of carriers result from their
interaction with one another (e.g. phase separation) and/or from
their interaction with an inhomogeneous distribution of dopants or
micro-structural defects such as dislocations \cite{39}. One can
assume that the electronic disorder or inhomogeneity in the cuprates
leads to the charge segregation into carrier-rich and carrier-poor
regions. As the doping level increases towards underdoped region,
specific charge ordering takes place in these regions and distinctly
different superlattices and energy bands of dopants (impurities with
trapped free carriers or large polarons) and self-trapped carriers
(intrinsic large polarons) are formed at their inhomogeneous spatial
distribution. In particular, the hydrogenic impurity centers
(impurities with loosely bound free carriers or large polarons) and
non-hydrogenic ones (impurities with tightly bound large polarons)
are assumed to form the superlattices with the lattice constant
$a_I$ and coordination number $z$. The charge ordering in
carrier-poor and carrier-rich domains results in the formation of
simple cubic, body-centered cubic and face-centered cubic
superlattices with coordination numbers $z=$6, 8 and 12,
respectively. In this case the formation of different impurity bands
in the cuprates is described by the tight-binding approximation and
the widths of the impurity bands can be determined from the relation
\begin{equation}\label{6Eq}
W_I=2zJ_I,
\end{equation}
where $J_I=\hbar^{2}/2m_{I}^{*}a_{I}^{2}$ is the hopping integral
between nearest-neighbour impurity centers, $m_{I}^{*}$ is the
effective mass of charge carriers in the impurity band.

Further, we believe that the intrinsic large polarons just like
impurity centers form different superlattices with the lattice
constant $a_p$ and the widths of the polaronic bands just as the
widths of the small-polaron bands \cite{40} are determined from the
expression
\begin{equation}\label{7Eq}
W_{p}=2zJ_{p},
\end{equation}
where $J_{p}=\hbar^{2}/2m_{p}^{*}a_{p}^{2}$ is the hopping integral
between nearest-neighbour sites of the polaronic superlattice,
$m_{p}^{*}$ is the effective mass of large polarons.

At low doping level the extrinsic (or intrinsic) large polarons form
a simple cubic superlattice with $a_I$ or $a_p>2R_p$, where $R_p$ is
the radius of the polaron. The width of the impurity band decreases
with increasing inter-dopant separation, so that the impurity-band
effective mass $m_I$ increases with inter-dopant separation or with
decreasing doping level. Clearly, in the lightly doped regime the
narrow impurity and/or polaronic bands are formed in the CT gap of
the parent cuprates and the energy gaps exist between the oxygen
valence band and the impurity or polaronic bands. These energy gaps
(or pseudogaps) disappear in sequence when the polaronic effects
disappear in carrier-rich and carrier-poor domains in the overdoped
regime.

We now make some remarks about the confinement energy of a large
polaron and the width of the large-polaron band in the cuprates. The
large-polaron confinement energy is about $\hbar^{2}/m_pR_p^{2}$
\cite{39,41}. When the Fr\"{o}hlich electron-phonon coupling
constant $\alpha$ is large ($\alpha\gg1$) the effective mass of a
large polaron is given by the formula \cite{40}
\begin{equation}\label{8Eq}
m_{p}=\frac{16m^{*}}{81\pi^{4}}\alpha^{4}
\end{equation}

The Fr\"{o}hlich coupling constant $\alpha$ for LSCO is estimated to
be $\alpha=5.7$ \cite{39,42}. For $m^{*}=m_e$ \cite{9}, we then
obtain from (8), $m_p\simeq2.14m_e$, which is close to the measured
value of $m_p=2m_e$ in LSCO \cite{9}. Using the Bohr radius of the
polaron $R_p=0.529\varepsilon_0(m_e/m_p){\AA}\simeq7.42{\AA}$ (at
$\varepsilon_0=30$) we obtain $\hbar^{2}/m_pR_p^{2}\simeq0.065eV$,
which is less than $E_p$ (see Sec. IV) and is larger than the
characteristic phonon energies $\hbar\omega=0.04-0.06 eV$ in the
cuprates \cite{9,23,43}. As the doping level is increased to
underdoped levels at which the insulator-to-metal transition takes
place, the distance between large polarons decreases and they can
form closely packed simple cubic superlattice with $a_p=2R_p$. In
this case the large-polaron bandwidth is about $W_p\simeq0.1 eV$
(i.e., $W_p\gtrsim E_p$). In contrast, the small-polaron band will
be very narrow ($W_p\ll\hbar\omega$) since the effective mass of a
small polaron is much larger than a free-electron mass.
\section{Metal-insulator transitions and stripe formation}\label{sec:level4}
Now, we discuss how the extrinsic and/or intrinsic self-trapping of
carriers, the dopant- and carrier-driven charge inhomogeneities and
the charge ordering are related to the MITs and stripe formation in
hole-doped cuprates. Various experiments have confirmed that for
light doping, $\rm{La_{2}CuO_{4}}$ behaves like a conventional
semiconductor \cite{9}. As described above, large-radius dopants or
impurities with $E_{dD}<0$ and $b_{s}<0$ may form the hydrogenic
acceptor centers in $\rm{La}$-based cuprates and the Hubbard model
based on the strong on-center Coulomb repulsion $U$ is applicable
for studying the MIT in these systems. Then, the Mott MIT point is
determined from the condition $W_{I}=E_{I}=1.16U$ which was used in
the derivation of the following criteria for the Mott MITs
\cite{19}: $n_{c}^{1/3}a_{H}\simeq0.289$ $(z=6)$,
$n_{c}^{1/3}a_{H}\simeq0.315$ $(z=8)$ and
$n_{c}^{1/3}a_{H}\simeq0.324$ $(z=12)$, where $n_c$ is the critical
concentration of the hydrogenic impurities initiating the Mott MIT
in doped systems. If we take $m^{*}=m_e$ (for free carriers) and
$\varepsilon_{0}=30$, we find $a_H=15.87{\AA}$ in LSCO. According to
the above Mott criteria, the MITs would occur at the dimensionless
hole concentrations $x_{c}=n_{c}/n_{a}\simeq0.00115-0.00161$, where
$n_{a}=1/V_{a}$ is the density of the host lattice atoms,
$V_{a}\simeq190 A^{3}$ is the volume per $\rm{CuO_2}$ unit in the
orthorhombic LSCO. The above values of $x_{c}$ are much smaller than
the value of $x_c\simeq0.02$ at which the destruction of the
antiferromagnetic (AF) order is observed in LSCO \cite{9}. However,
the large-radius dopant and large polaron may form the hydrogenic
impurity center in LSCO. When $b_{s}<0$ (or $E_{dD}<0$), the defect
and the self-trapped hole repel each other at short distance but
attract each other at long distance. Therefore, the hydrogenic
self-trapped state of a large polaron in LSCO has an effective Bohr
radius $a_{H}=0.529\varepsilon_{0}(m_{e}/m_{p}){\AA}$. The existence
of such hydrogenic acceptor centers in $\rm{La}$-based cuprates is
now experimentally well established \cite{9}.

According to Mott and Davis \cite{38}, the Mott's criterion for the
MIT can be used as the condition for the metallic behavior of a
degenerate polaron gas if the Bohr radius $a_H$ of polarons is
larger than the interatomic distance. If we take $m_{p}=2m_{e}$ and
$\varepsilon_{0}=27$, we find $a_{H}\simeq7.14{\AA}$, so that the
above Mott criteria for the MITs can be applied to the cuprates. In
this case the Mott transitions would occur at $x_c$=0.0112-0.0178.
The quantitative criteria for the Anderson MITs,
$n_{c}^{1/3}a_{H}\simeq0.289$ (for $z$=6),
$n_{c}^{1/3}a_{H}\simeq0.293$ (for $z$=8), and
$n_{c}^{1/3}a_{H}\simeq0.272$ (for $z$=12), derived in Ref.
\cite{19} predict nearly the same values of $x_c$. We see that the
Mott transition (for shallow impurity centers) would occur at a much
smaller dopant concentration than the critical doping concentration
for the MIT in the cuprates. For example, the MITs in LSCO are
observed at more higher doping levels $x_{c}\simeq0.05-0.07$
\cite{2,3,10,16,25} and these experimental results cannot be
reconciled with the above criteria for the Mott MITs. In our
opinion, the hole carriers liberated from the large-radius dopants
or hydrogenic impurity centers at $x_{c}\gtrsim0.02$ are constrained
to remain away from these impurities and they are self-trapped in a
defect-free deformable lattice with the formation of strong-coupling
intrinsic large polarons, which can acquire itineracy at higher
doping levels \cite{28}.

Another possibility is that small-radius dopants (e.g.,
$\rm{Ca^{2+}}$ and $\rm{Nd^{3+}}$) with $E_{dD}>0$ (or $b_{s}>0$)
favor the formation of the extrinsic large polarons (non-hydrogenic
impurity centers) in $\rm{La}$-based cuprates. In the present case,
both the short and long range parts of the carrier-defect
interaction is attractive, so that hole carriers are self-trapped
near such dopants. Polaronic effects will be stronger near the
small-radius dopants, leading to carrier localization over a broader
range of doping. We can evaluate the possibility for the existence
of a potential barrier separating large- and small-radius intrinsic
polaronic states in 3D hole-doped $\rm{La}$-based cuprates using the
relation (\ref{4Eq}).

According to the spectroscopic data, the Fermi energy of the undoped
cuprates is about $E_{F}\simeq7 eV$ \cite{44,45}. To determine the
value of the short-range carrier-phonon coupling constant $g_{1}$,
we estimate $E_{d}$ as $E_{d}=(2/3)E_F$. The values of other
parameters are $K=1.4\cdot10^{12}dyn/cm^2$ \cite{46}, $m^{*}=m_{e}$
\cite{9}, $\varepsilon_{\infty}=3-5$ \cite{1}, $a_{0}\simeq6{\AA}$
(for orthorhombic LSCO), $b_{s}=0$ and $Z=0$. Then $B=1$ eV,
$g_{1}\simeq0.0576$ and
$g_{2}\simeq2.4(1-\eta)/\varepsilon_{\infty}$.  For the cuprates,
typical values of $\eta$ range from 0.02 to 0.12 \cite{9}. Using
Eq.(\ref{4Eq}), we obtain $E_a$ value of $\sim38$ eV at
$\varepsilon_{\infty}=4$ and $\eta=0.06$. As already seen, the
large- and small-radius polaronic states are separated by very high
potential barrier. This barrier prevents the formation of small
polarons in the 3D cuprates. At $b_{s}=0.5$ and $Z=1$ the height of
such a potential barrier separating large- and small-radius
extrinsic polaronic states is $\sim14$ eV. It follows that the
relevant charge carriers in the anisotropic 3D cuprates are the
large intrinsic and extrinsic polarons (or bipolarons). While the
small polarons and bipolarons may be formed in the $\rm{CuO_2}$
layers of the cuprates, where the self-trapping of carriers may
occur more easily due to the absence of the potential barrier
between the large- and small-radius self-trapped states in 2D
systems \cite{47,48}. However, such (bi)polarons having very large
effective masses and narrow energy bands tend to be localized rather
than mobile.

Small or large polarons may be expected in polar materials with a
small or large hopping integral $J$ between nearest-neighbour sites
of the crystal lattice. The radius of the polaron $R_p$ is of order
$a_{0}(2J/E_{p})$ and the formation of a polaron whose radius is
large compared to the lattice constant $a_0$ requires that
$2J/E_{p}\gg1$ (see also Ref. \cite{49}) or $W_p/E_p>1$ \cite{50}.
In hole-doped cuprates (LSCO, $\rm{YBa_2Cu_3O_{7-\delta}}$ and
$\rm{Bi_2Sr_2CaCu_2O_{8+\delta}}$) large polarons have the effective
masses $m_{p}\simeq(2\div4)m_e$ \cite{9,51}. Therefore, $W_I$ and
$W_p$ may be comparable with the binding energies $E_{p}^{I}$ and
$E_{p}$ of the extrinsic and intrinsic large polarons since $m_p$,
$m_I(\gtrsim m_p$), $a_I$ and $a_p$ would decrease with increasing
carrier concentration. At $m_I\lesssim2m_e$ and $m_{p}\lesssim 2m_e$
the magnitudes of $W_I$ and $W_p$ may become larger than $E_{p}^{I}$
and $E_{p}$, which are determined from Eq.(\ref{5Eq}) by using the
above presented values of parameters. The obtained results are
summarized in Table 1. When $\eta$ increases, the ionization energy
$E_{p}^{I}$ of the non-hydrogenic impurity centers (with $Z=1$)
increases. On the contrary, the binding energy of the intrinsic
large polarons $E_p$ decreases. Interestingly, $E_{p}^{I}$ and
$E_{p}$ increases markedly with decreasing $\varepsilon_{\infty}$
from 5 to 3. We see that the value of $E_{p}^{I}$=0.13 eV obtained
at $\eta$=0.10 and $Z$=1 is consistent with the experimental data
for lightly doped $\rm{La_2CuO_{4+y}}$ \cite{9} and the impurity
band observed in these systems at 0.13 eV is associated with the
extrinsic large polarons rather than the small polarons bound to
impurities.
\begin{table}
\caption{\label{tab:table1} Binding energies of intrinsic and
extrinsic large polarons for $\varepsilon_{\infty}=4$, $b_{s}=0.5$
and different values of $\eta$.}
\begin{ruledtabular}
\begin{tabular}{ccccccccccc}
$\eta$      & 0.00  & 0.02  & 0.04  & 0.06  & 0.08  & 0.10  & 0.12\\
\hline
$E_{p}$, eV  & 0.0916 & 0.0879 & 0.0844 & 0.0809 & 0.0774 & 0.0741 & 0.0708\\
\hline &&&$Z=0$&&&\\
\hline $E_{p}^{I}$, eV & 0.0925 & 0.0888 & 0.0851 & 0.0816 &0.0781 & 0.0747 & 0.0714\\
\hline &&&$Z=1$&&&\\
\hline
$E_{p}^{I}$, eV & 0.0925 & 0.0995 & 0.1067 & 0.1142 & 0.1220 & 0.1301 & 0.1384\\
\end{tabular}
\end{ruledtabular}
\end{table}
\par
As we have discussed above, the extrinsic large polarons bound to
dopants may also form different superlattices in inhomogeneous
hole-doped cuprates and the energy bands (which are formed in the
lightly doped cuprates) of such polarons may exist, thus permitting
charge transport by means of intra-band conduction. In case of
narrow polaronic bands, charge transport becomes hopping-like and is
caused by intra-band hopping processes \cite{52}. One can assume
that if the bandwidth of the extrinsic large polarons exceeds some
critical value, their intra-band conduction becomes metal-like. We
attempt to find the new criteria for such MITs in doped materials.

When the carrier concentration is low, the carriers occupy low-lying
bound states (i.e., extrinsic or intrinsic polaronic states) first.
As soon as all the localized states were filled at some critical
carrier concentration $n=n_{c}$ (or $x=x_{c}$), the carriers start
to occupy the itinerant states and the transition of the system from
the insulating phase to the metallic one occurs. The possible values
of $n_{c}$ at which extrinsic or intrinsic large polarons forming
different superlattices acquire itineracy are determined from the
appropriate criteria for the MITs. The conditions for carrier
localization or delocalization can be obtained by using the
uncertainty principle: $\Delta p\Delta x\geq\hbar/2$, where $\Delta
p$ and $\Delta x$ are the uncertainties in the momentum and
coordinate, respectively. This uncertainty relation can be written
as
\begin{equation}\label{9Eq}
\Delta x\cdot\Delta E\simeq \frac{\hbar^{2}(\Delta
k)^{2}}{2m^{\ast}}\cdot\frac{1}{2\Delta k},
\end{equation}
where $\Delta E$ and $\Delta k$ are the uncertainties in the energy
and wave vector, respectively. The expression $\hbar^{2}(\Delta
k)^{2}/2m^{\ast}$ in Eq. (\ref{9Eq}) represents the uncertainty in
the energy of free carriers. Taking into account that the
uncertainties in the energy and wave vector in the impurity band are
about $W_{I}$ and $1/a_{I}$, respectively, the relation (\ref{9Eq})
can be rewritten in the form (cf. a similar relation formerly
derived by Ridley \cite{41} using the uncertainty principle)
\begin{equation}\label{10Eq}
\Delta x\cdot\Delta E\cong W_{I}a_{I}/2
\end{equation}
On the other hand, the uncertainty in the energy $\Delta E$ of the
localized state of carriers is of the order of $E_{p}^{I}$, whereas
the uncertainty in the coordinate $\Delta x$ of the extrinsic large
polarons is of the order of the polaron radius $R_{I}$. Then, the
condition for the setting-in of carrier localization can be written
as
\begin{equation}\label{11Eq}
E_{p}^{I}R_{I}\gtrsim W_{I}a_{I}/2,
\end{equation}
so that the new criterion for the MIT can be derived from this
condition and written as
\begin{equation}\label{12Eq}
\frac{E_{p}^{I}}{W_I}=0.5\frac{a_I}{R_I},
\end{equation}
which is similar to the results for analogous problems of Holstein
localization, Mott and Anderson MITs. In particular, localization in
the one-dimensional Holstein molecular-crystal model is governed by
the ratio of the local polaron binding energy to the inter-site
bandwidth \cite{39,53}. Similarly, Anderson's localization in
Anderson's disorder model is governed by the ratio of the local-site
energy (or the random potential $V_0$) to the electron bandwidth and
Mott's MIT is governed by the ratio of the Hubbard on-site Coulomb
interaction energy $U$ to the $W$ \cite{38,41,52}. For intrinsic
large polarons, instead of (\ref{12Eq}), we could write
\begin{equation}\label{13Eq}
\frac{E_{p}}{W_p}=0.5\frac{a_p}{R_p}.
\end{equation}

With increasing the inter-polaron distance $R$, the polaron band is
continuously narrowed, finally ending in the discrete levels (at
$R\gg R_p$) of the non-interacting polarons. On the contrary, with
decreasing $R$, the large polarons begin to interact one another
through the overlap of their structural distortions and the
polaronic states are  broadened into an energy band.  As long as the
distance between nearest-neighbour polarons is larger than $2R_{p}$
and the overlap of the deformation clouds of polarons is small, one
would expect the large-polaron bands to be very narrow. In this case
large polarons are localized or confined to their potential wells
and the system is converted into an insulator. For $R=2R_p$ (which
is assumed to correspond to the packing of spheres with radii equal
to the polaron radius), the deformation clouds of large polarons
begin to overlap strongly and the large polarons are delocalized and
characterized by metal-like transport in sufficiently broadened
energy bands.

If the extrinsic large polarons form such closely packed simple
cubic, body-centered cubic and face-centered cubic superlattices
with $a_{I}=2R_{I}$ ($z=6$), $a_{I}=(4/\sqrt{3})R_{I}$ ($z=8$) and
$a_{I}=(2\sqrt{2})R_{I}$ ($z=12$), the appropriate densities of such
carriers per unit cells of the superlattices are $n=1/a_{I}^{3}$,
$n=2/a_{I}^{3}$ and $n=4/a_{I}^{3}$, respectively. For these cases,
three successive MITs are feasible and the localized extrinsic large
polarons in the cuprates start to occupy the itinerant states at
different hole concentrations determined from relation (\ref{12Eq}).
This means that the system gradually changes from the insulating
phase to the metallic one. When the extrinsic large polarons form
simple cubic and face-centered cubic superlattices, such
insulator-to-metal transitions occur at $W_{I}=E_{p}^{I}$ and
$W_{I}=E_{p}^{I}/\sqrt{2}$, respectively. We argue that the
electronic inhomogeneities and the ordering of polaronic carriers
with the formation of different superlattices can produce various
types of self-organized electronic structures in the form of
stripes. Using the relation (\ref{12Eq}), we obtain the following
criteria for the new MITs:
\begin{equation}\label{14Eq}
n_{c}=\left(\frac{m_{I}E_{p}^{I}}{z\hbar^{2}}\right)^{3/2}\quad
\textrm{for} \quad z=6,
\end{equation}
\begin{equation}\label{15Eq}
n_{c}=\frac{1}{\sqrt{2}}\left(\frac{\sqrt{3}m_{I}E_{p}^{I}}{z\hbar^{2}}\right)^{3/2}\quad
\textrm{for} \quad z=8,
\end{equation}
\begin{equation}\label{16Eq}
n_{c}=2^{5/4}\left(\frac{2m_{I}E_{p}^{I}}{z\hbar^{2}}\right)^{3/2}\quad
\textrm{for} \quad z=12.
\end{equation}

We are now in a position to evaluate $n_c$ in $\rm{La}$-based
cuprates using Eqs. (\ref{14Eq}), (\ref{15Eq}) and (\ref{16Eq}). The
effective masses of carriers in cuprates deduced from ARPES and
electronic specific heat data (see Ref. \cite{54}) at different
doping levels are slightly different and equal to 2.1 - 2.5 times
the free electron mass. Therefore, we can evaluate $n_c$ by taking
$m_{I}=2.5m_e$ for $\rm{La_{2-x}Ca_{x}CuO_{4}}$. Then for
$\eta=0.02$ and $E_{p}^{I}=0.10 eV$ we find
$x_{c}\simeq0.065-0.080$. When $\varepsilon_{\infty}=4$,
$\eta=0.02-0.12$ and $E_{p}^{I}=0.100-0.138$ eV (see Table 1),  the
MITs and stripe formation occur in these systems at
$x_{c}\simeq0.065-0.131$. At $\varepsilon_{\infty}=3$ and
$\eta=0.06$ we obtain $E_{p}^{I}\simeq0.2eV$. Then the
metal-insulator boundary of $\rm{La_{2-x}Ca_{x}CuO_{4}}$ lies in the
overdoped regime at $x_c\simeq0.226$ (for $z=8$).

Further, the double substitution of smaller cations for host lattice
ions and dopants may also favor the MITs and stripe formation
occurring in a wide range of doping of $\rm{La}$-based cuprates.
When the doping increases, there is a significant probability of at
least two neighbouring $\rm{La^{2+}}$ ions replaced by large-radius
$\rm{Sr^{2+}}$ ion (with $Z=1$, $V_{0}^{Sr}>0$ and $E_{dD}^{Sr}<0)$
and by small-radius $\rm{Nd^{3+}}$ ion (with $Z=0$, $V_0^{Nd}>0$ and
$E_{dD}^{Nd}>0$) in $\rm{La_{2-x-y} Nd_{y}Sr_{x}CuO_{4}}$ (where
$y>>x$). The attractive potentials of these two-dopant centers may
be greater than those of more separated dopants. For
$\rm{La_{2-x-y}Nd_{y}Sr_{x}CuO_{4}}$ with such two-dopant centers,
we should replace $V_0$ and $E_{dD}$ by $V_{0}^{Sr}+V_{0}^{Nd}$ and
$E_{dD}^{Sr}+E_{dD}^{Nd}$ in Eq.(\ref{1Eq}) or in the expression for
$b_s$. At $E_{dD}^{Sr}+E_{dD}^{Nd}>0$ (i.e., $b_{s}>0$) the
short-range part of the two-dopant potential is attractive and the
hole carriers are self-trapped near the two neighbouring dopants
with the formation of the extrinsic large polarons. The
two-dopant-driven charge inhomogeneity and ordering lead to
formation of the superlattices of such extrinsic large polarons. In
order to illustrate the effects of anisotropy of
$\varepsilon_\infty$ and the short-range defect potential on $x_c$,
we show in Fig.1 results of our calculations for hole-doped
$\rm{La}$-based cuprates with two-dopant centers for which $Z=1$ and
$b_{s}>0$. Fig.1 shows that the anisotropy of $\varepsilon_\infty$
and the size effect of small-radius dopants are the main driving
forces for the new MITs and stripe formation in a wide range of
doping (including also $x=1/8$) in $\rm{La}$-based cuprates
containing two types of dopants (i.e., large- and small-radius
dopants). Indeed, in hole-doped cuprates
$\rm{La_{2-x-y}Nd_{y}Sr_{x}CuO_{4}}$ and
$\rm{La_{2-x-y}Eu_{y}Sr_{x}CuO_{4}}$ with small-radius dopants
$\rm{Nd^{3+}}$ and $\rm{Eu^{3+}}$, the static stripe phases are
observed in a wide range of carrier concentration and not restricted
to a narrow range around $x=1/8$ in which superconductivity is
suppressed \cite{16,27,55}.
\begin{figure}
\includegraphics[width=0.5\textwidth]{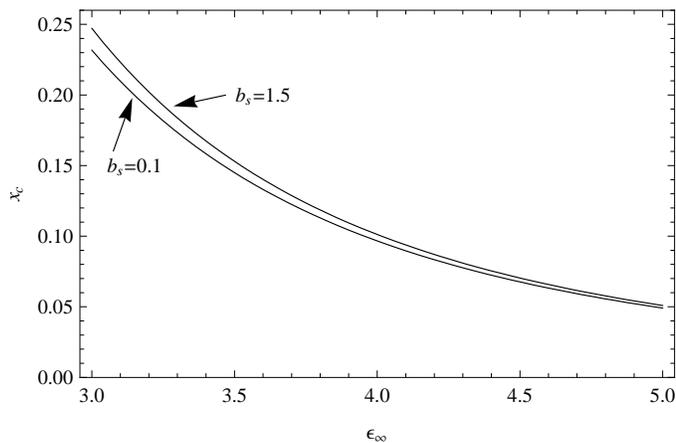}
\caption{\footnotesize Critical carrier concentration $x_{c}$ as a
function of $\varepsilon_{\infty}$ at $z=8$, $\eta=0.06$,
$b_{s}=0.1$ and $b_{s}=1.5$.}
\end{figure}

As mentioned above, the size of impurities is also a key parameter
in the cuprates containing large-radius dopants and the formation of
intrinsic large polarons in these systems becomes possible at
distances far from the impurities. Such materials are the compounds
LSCO and $\rm{La_{2-x}Ba_{x}CuO_{4}}$ (LBCO) in which large polarons
form different superlattices and acquire itineracy at some critical
carrier concentrations $n_c$ determined from the relations
(\ref{14Eq}), (\ref{15Eq}) and (\ref{16Eq}) by replacing $m_I$ and
$E^{I}_{p}$ by $m_p$ and $E_{p}$ in these relations. Using the
values of $\eta=0.02$, $\varepsilon_{\infty}=4$, $m_{p}=2.1m_{e}$
and $E_{p}\simeq0.088 eV$ (see Table 1) for LSCO and LBCO, we find
$x_c\simeq0.049$ (for $z$=6) and $x_c\simeq0.051$ (for $z$=8). These
values of $x_c$ are in excellent agreement with the experimental
value $x_c=0.05$ in LSCO \cite{10}. If we take into account a
possible anisotropy of $\varepsilon_{\infty}=3-4$ and
$\eta=0.02-0.10$, then we obtain $x_c\simeq0.04-0.125$, which are
also well consistent with existing experimental data on MITs  and
stripe formation in LCSO and LBCO (see Refs. \cite{10,22,26,55}). In
Fig. 2 we plot the dependence of $x_c$ on $\eta$ for $z=8$,
$b_s=0.5$ and $\varepsilon_{\infty}=4$.
\begin{figure}
\includegraphics[width=0.5\textwidth]{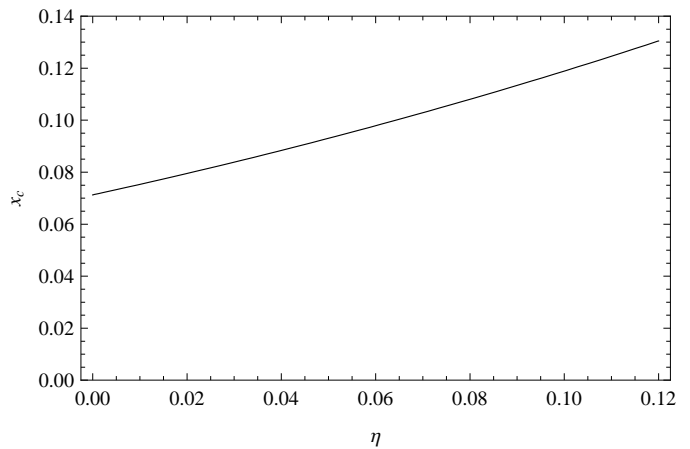}
\caption{\footnotesize Critical carrier concentration $x_{c}$ as a
function of  $\eta$ at $z=8$, $b_{s}=0.5$ and
$\varepsilon_{\infty}=4$.}
\end{figure}
Fig.2 shows that the value of $x_c$ increases markedly with
increasing $\eta$ for the cuprates containing large-radius dopants.
Our results clearly indicate that $x=1/8$ is truly "magic" or
particular doping level for hole-doped cuprates LSCO and LBCO. In
these systems, the suppression of superconductivity should be
expected for $x$ close to $1/8$. However, the doping level $x=1/8$
is not "magic" doping level for the formation of stripes in other
hole-doped cuprates, where the effects of small-radius dopants on
MITs and static stripe formation are strong enough. The stripe
formation driven by the MITs occurs in such systems not only at
$x=1/8$ but also at much more higher doping levels. Indeed,
experimental studies on $\rm{La_{2-x}Ba_{x-y}M_{y}CuO_{4}}$ (where
$\rm{M=Sr,Ca}$) show that the 1/8 anomaly is reduced by substituting
smaller divalent cations instead of large-radius $\rm{Ba^{2+}}$
\cite{55}. In contrast, the ordering of intrinsic large polarons is
important for the MITs and stripe formation in LSCO and LBCO. Double
doping experiments on $\rm{La_{2-x-y}Ba_{x}Th_{y}CuO_{4}}$ indicate
that superconductivity is suppressed when the hole concentration
$p=x-y$ (not the individual or total dopant concentrations $x$, $y$
and $x+y$) is 1/8 \cite{56,57}. In this system, the $\rm{Th^{4+}}$
ion acts to compensate the negative effective charge of
$\rm{Ba}$-site, so that the hole concentration is now given by
$p=x-y$. This means that the presumed superlattice and stripe
formation accompanied by the suppression of superconductivity in
$\rm{La_{2-x-y}Ba_{x}Th_{y}CuO_{4}}$ are associated with the
ordering of intrinsic large polarons \cite{39} rather than dopants.

The formation of 3D and 2D large bipolarons is also expected in the
cuprates \cite{1,5,6,58,59,60}. Such bipolarons can only exist if
the Fr\"{o}hlich coupling constant $\alpha$ is greater than the
critical value $\alpha_{c}$ and when $\eta$ is smaller than the
critical value $\eta_{c}$. For instance, for 3D and 2D systems, the
following critical parameters were obtained by using the Gaussian
type trial wave functions: $\alpha_{c}^{(3D)}\simeq5.8$,
$\eta_{c}^{(3D)}\simeq0.13$ \cite{6,59} and
$\alpha_{c}^{(2D)}\simeq2.94$, $\eta_{c}^{(2D)}\simeq0.167$ (for
$\alpha=5$), $\eta_{c}^{(2D)}\simeq0.173$ (for $\alpha=10$)
\cite{60}. The ratio of 3D bipolaron binding energy $E_{bB}$ to
twice the polaron energy $2E_{p}$ is about $E_{bB}/2E_{p}\simeq0.22$
at $\eta\rightarrow0$ \cite{61}, whereas the ratio of 3D extrinsic
bipolaron binding energy $E_{bB}^{I}$ to twice the extrinsic polaron
energy $2E_{p}^{I}$ is somewhat smaller than the value of
$E_{bB}/2E_{p}$ \cite{62}. Furthermore, the binding energies of 3D
extrinsic and intrinsic large bipolarons decrease with the increase
of $\eta$. Therefore, such bipolarons may exist at low doping
levels. As the carrier concentration increases, 3D large bipolarons
dissociate first into two large polarons and then these polarons
acquire itineracy at a higher doping level $x=x_c$ which corresponds
to the onset of the metallic behavior of the system. However, the
properties of 2D large (bi)polarons may be quite different from
those of 3D large (bi)polarons. It is clear that the binding energy
and the effective mass of a bipolaron would increase greatly with
decreasing dimensionality. In particular, the binding energy of 2D
large bipolarons $E_{bB}^{(2D)}$ is expected to be much greater than
$E_{bB}^{(3D)}$. The 2D polaron and bipolaron ground-state energies
$E_{p}^{(2D)}$ and $E_{bB}^{(2D)}$ can be obtained approximately
using the scaling relations \cite{58,63}:
\begin{eqnarray}
E_{p}^{(2D)}(\alpha)=\frac{2}{3}E_{p}^{(3D)}((3\pi/4)\alpha)
\end{eqnarray}
and
\begin{eqnarray}
E_{B}^{(2D)}(\alpha,U)=\frac{2}{3}E_{B}^{(3D)}\left(\frac{3\pi}{4}\alpha,\frac{3\pi}{4}U\right),
\end{eqnarray}
where $E_{p}^{(3D)}$ and $E_{B}^{(3D)}$  are the 3D polaron and
bipolaron ground-state energies, respectively, U is  the strength of
the Coulomb repulsion between the pairing carriers.

In the strong coupling regime $\alpha>\alpha_{c}$, the ground-state
energies of 3D large polaron and bipolaron are proportional to
$\alpha^{2}$  \cite{59}. In this case the 2D large bipolaron binding
energy is given by
\begin{eqnarray}
E_{bB}^{(2D)}=\frac{2}{3}\left(\frac{3\pi}{4}\right)^{2}E_{bB}^{(3D)}\simeq3.7E_{bB}^{(3D)}
\end{eqnarray}
from which it follows that the bipolaron binding energy will be much
larger in 2D systems as compared to 3D ones and, therefore, 2D
bipolarons will be strongly localized. Such 2D large bipolarons will
unlikely be dissociated in underdoped and optimally doped cuprates.
According to Emin \cite{5,39,64,65}, quasi-2D large bipolarons are
attracted to one another via the phonon-mediated interaction that
fosters their condensation into a normal liquid. Carriers are
assumed to condense into droplets and form a liquid prior to their
condensing further into a superconducting state \cite{39}. The
concentration of such large bipolarons above which they condense
into a normal liquid at a reasonable temperature is assumed to be
$x_{bi}^{crit}\sim0.1/n_{bi}$ \cite{5,39}, where $n_{bi}$ is the
number of unit cells involved in a large bipolaron. To form one
bipolaron, two carriers are required from two dopants, then the
critical doping level is $x_{c}=2x_{bi}^{crit}$. Envisioning a large
bipolaron involving about five cells, one cell and its four
neighbors, the critical carrier concentration for forming a
large-bipolaron liquid is about $x_{c}\simeq0.2/5=0.04$ \cite{5,39}.
It was also suggested \cite{5,39} that in the cuprates,
superconductivity  occurs when the dopant density exceeds the
density of a large bipolaron liquid, namely, at $x_{c}\gtrsim0.05$,
and large-bipolaron superconductivity of the cuprates is determined
by the competition between large-bipolarons' attraction for one
another and their attraction to dopants. At the same time the
condensation of attracting large bipolarons (including also Cooper
pairs of large polarons) into a superfluid Bose-liquid in the
cuprates below the superconducting transition temperature $T_{c}$
was suggested in Refs. \cite{66,67}. The condensation of large
bipolarons into a superfluid Bose-liquid at sufficiently low
temperatures in the cuprates can be regarded as an
insulator-to-superconductor transition \cite{28}. Emin argued
\cite{5,39,64,65} that (bi)polarons in the cuprates order in a
manner commensurate with the underlying lattice, thereby forming a
superlattice. For the square structure of $\rm{CuO_{2}}$ planes such
a commensurate ordering of large bipolarons can occur at $x=1/8$ and
lead to the conversion of a cuprate superconductor to an insulator
composed of ordered large bipolarons \cite{68}. Thus, the type of
transition proposed by Emin is another possibility of the
metal-to-insulator (at $T>T_{c}$) or superconductor-to-insulator (at
$T\lesssim T_{c}$) transition which is driven by the commensurate
ordering of quasi-2D large bipolarons and presumed superlattice
formation. It is believed \cite{39} that such transitions are
neither a Mott transition nor an Anderson transition.

\section{Conclusions}\label{sec:level5}

In this paper we have studied the carrier localization, MITs and
stripe formation in inhomogeneous hole-doped cuprates driven by the
gross electronic inhomogeneities, charge ordering and extrinsic
and/or intrinsic self-trapping of hole carriers. To model the real
situation in cuprates, the hole carriers are presumed to exist in a
3D ionic continuum medium and interact with the dopants (or
impurities) via short and long-range carrier-defect-lattice
interactions and with the lattice vibrations via short and
long-range carrier-lattice interactions. We have investigated the
self-trapping of carriers near the defects and away from the defects
(i.e., in a defect-free deformable lattice) within the continuum
model and adiabatic approximation. Then, the possibility of the
formation of extrinsic self-trapped states (i.e., hydrogenic and
non-hydrogenic acceptor states) and intrinsic ones (large-polaronic
states) in the CT gap of the cuprates is examined. We think that the
large-radius dopants with the short-range repulsive potential can
form hydrogenic impurity centers in LSCO and LBCO, where the hole
carriers liberated from such impurity centers are self-trapped in a
defect-free deformable lattice due to the strong short and
long-range carrier-lattice interactions. When such hydrogenic
impurity centers are distributed inhomogeneously and ordered
differently with the formation of distinctly different
superlattices, the narrow Hubbard impurity bands are formed in the
CT gap of the cuprates. We have analyzed the applicability of the
newly derived criteria for the Mott and Anderson MITs to the
$\rm{La}$-based cuprates. It is found that such MITs in the cuprates
with the hydrogenic impurity centers are unlikely possible. The
situation in $\rm{La}$-based cuprates with small-radius dopants is
quite different from LSCO and LBCO cases. In these systems, the
short- and long-range parts of the small-radius dopant potential are
attractive and the combined effect of the strong carrier-defect and
carrier-lattice interactions fosters carrier localization (or
defect-assisted self-trapping of carriers) and formation of
extrinsic large polarons. We have calculated the ground state
energies (i.e., binding energies) of extrinsic and intrinsic large
polarons in $\rm{La}$-based cuprates. The obtained results are
consistent with experimental results on polaron formation in these
systems.

Further, we have demonstrated  that the driving forces for carrier
localization, MITs and stripe formation in $\rm{La}$-based cuprates
are the extrinsic and/or intrinsic self-trapping of hole carriers,
the dopant and carrier-driven electronic inhomogeneities, the charge
ordering and the anisotropy of dielectric constants. Under the
certain conditions, the narrow energy bands of extrinsic and
intrinsic large polarons are assumed to prelude itinerant motion of
such polarons. These conditions and quantitative criteria for the
new MITs in inhomogeneous $\rm{La}$-based cuprates are obtained by
using the pertinent uncertainty relation. We showed that the new
MITs are accompanied by the formation of static and dynamic stripes,
and would occur at different critical carrier concentrations
depending on the ionic size of dopants (or impurities), the
anisotropy of dielectric constants and the type of the ordering of
extrinsic and/or intrinsic large polarons. In particular, the
ordering of 3D intrinsic large polarons, the formation of different
polaronic superlattices ($z=6$ and 8) and the anisotropy of
dielectric constants favor the MITs and stripe formation in the
range of doping $x=0.04-0.125$ in $La$-based cuprates with
large-radius dopants. We argue that the $x=1/8$ anomaly is
characteristic of these cuprates, where the suppression of
superconductivity due to the formation of static stripes should be
especially pronounced for the doping level $x=1/8$. While the
effects of the substitution of smaller cations for host lattice ions
or dopants (e.g., $\rm{Nd^{3+}}$ in LSCO), the anisotropy of
dielectric constants and the ordering of extrinsic large polarons
with the formation of different superlattices ($z=6$ and 8) extend
the metal-insulator crossover region from the heavily underdoped
($x=0.04$) to the overdoped ($x>0.2$) regime. Our results suggest
that the new MITs and the formation of static stripes (in
carrier-poor domains) and dynamic ones (in carrier-rich domains)
occur in a wide range of doping (including also the "magic" doping
1/8) in $La$-based cuprates containing small-radius dopants (e.g.,
$\rm{Ca^{2+}}$ and $\rm{Nd^{3+}}$ ions), where the hole
concentration $x=1/8$ is not already particular doping level
corresponding to the formation of static stripes.

When 3D extrinsic and intrinsic large bipolarons are formed in the
cuprates at $\eta<\eta_{c}$, such bipolarons will dissociate first
into two large polarons at low carrier concentration and then these
polarons become mobile or acquire itineracy at high doping levels
$x\geq x_{c}\simeq0.05$. The essence of the above driving forces for
carrier localization, MITs and static stripe formation in
$\rm{La}$-based cuprates has been mimicked by the theoretical
predictions which capture the main characteristics of experiment.
Thus, we have succeeded in explaining the experimental results on
carrier localization, MITs and stripe formation in hole-doped
$\rm{La}$-based cuprates with large- and small-radius dopants in the
two characteristic doping regimes $0.05\leq x\leq0.125$ and
$0.05\leq x\leq0.25$, respectively.

Finally, it should be noted that carrier localization and metal (or
superconductor)-to-insulator transition accompanied by the 1/8
anomaly in hole-doped cuprates might be also driven by the
commensurate ordering of quasi-2D (or planar) large bipolarons which
will condense into a liquid above some critical carrier
concentration.
\begin{center}
{\bf ACKNOWLEDGMENTS}
\end{center}
Over the years we have benefited through collaboration,
communication and discussions on the copper-oxide problem by D.
Emin. We thank D. Emin for his careful reading of and for comments
on the manuscript. Discussions with B. Yavidov and E.I. Ibragimova
on aspect of carrier self-trapping and metal-insulator transition
phenomena are also gratefully acknowledged. This work was supported
by the STCU grant 3505 and the Foundation of the Fundamental
Research of Uzbek Academy of Sciences, Grant FA-F2-F070.

\bibliography{apssamp}

\begin{thebibliography}{00}\label{sec:TeXbooks}
\bibitem{1}D. Emin and M.S. Hillery, Phys. Rev. B39, 6575 (1989)
\bibitem{2} J.Fink, N. N\"{u}cker, M.Alexander, H. Romberg, M. Knupeer, M.
Merkel, P.Adelmann, R. Claessen, G. Mante, T.Buslaps,
S.Harm,R.Manzke and M. Skibowski, Physica C 185-189, 45 (1991)
\bibitem{3} M. Cieplak, S. Guha, H. Kojima, P. Lindenfeld,
G. Xiao, J.Q. Xiao and C.L. Chien, Physica C 185-189, 1233 (1991)
\bibitem{4} K. Nasu, Physica C 185-189, 1595 (1991)
\bibitem{5} D. Emin, Phys.Rev.Lett. 72, 1052 (1994); Phys.Rev. B49, 9157
(1994); (private communication)
\bibitem{6} S. Dzhumanov, P.J. Baimatov, A.A. Baratov and P.K.
Khabibullaev, Physica C254, 311 (1995)
\bibitem{7} P.Quemerias, Mod.Phys.Lett. B 9, 1665 (1995)
\bibitem{8} G.S. Boebinger, Y. Ando, A. Passner, T. Kimura, M. Okuya,
J. Shimoyama, K. Kishio, K. Tamasaku, N. Ichikawa and S. Uchida,
Phys.Rev.Lett. 77, 5417 (1996)
\bibitem{9} M.A. Kastner, R.J. Birgeneau, G. Shirane and Y. Endoh, Rev.Mod.Phys. 70, 897 (1998)
\bibitem{10} M. Imada, A. Fujimori and Y.Tokura, Rev.Mod.Phys. 70, 1039 (1998)
\bibitem{11} C. Castellani, C. Di Castro and M. Grilli, J.Phys.Chem. Solids. 59,1694 (1998)
\bibitem{12} A.N. Lavrov and V.F. Gandmakher, Phys. Usp. 41, 223 (1998)
\bibitem{13} J. Zaanen, eprint, cond-mat/0103255
\bibitem{14} S.A. Kivelson, I. Bindloos, E. Fradkin, V. Oganesyan,
J. Tranquada, A. Kapitulnik and C. Howard, Rev.Mod.Phys. 75, 1201 (2003)
\bibitem{15} A.A. Abrikosov, Physica C 460-462, 1 (2007)
\bibitem{16} M. H\"{u}cker, G.D. Gu, J.M. Tranquada , M.V. Zimmerman, H.-H. Klauss,
N.J. Curro, M. Braden and B. B\"{u}chner, Physica C 460-462 170
(2007)
\bibitem{17} A.J. Millis, P.B. Littlewood and B.I. Shraiman, Phys. Rev. Lett. 74, 5144 (1995)
\bibitem{18} W.E. Pickett and D.J. Singh, Phys.Rev. B 55, R8642 (1997)
\bibitem{19} S. Dzhumanov, U.T. Kurbanov and A. Kurmantayev, Int. J. Mod. Phys. B21,169 (2007)
\bibitem{20} T. Kato, T. Noguchi, R. Saito, T. Machida, and H. Sakata, Physica C 460-462, 880 (2007)
\bibitem{21} K. Omori, T. Adachi, Y. Tanabe and Y. Koike, Physica C 460-462, 1184 (2007)
\bibitem{22} P.A. Lee, N. Nagoasa and X.-G. Wen, Rev. Mod. Phys. 78, 17 (2006)
\bibitem{23} S. Sugai, Physica C 185-189, 76 (1991)
\bibitem{24} D. Mihailovic, C.M. Foster, K. Voss and A.J. Heeger, Phys. Rev. B42, 7989 (1990)
\bibitem{25} Z. Konstantinovic, Z.Z. Li and H. Raffy, Physica C 351, 163 (2001)
\bibitem{26} S. Ono, Y. Ando, T. Murayama, F.F. Balakirev, J.B.
Betts and G.S. Boebinger, Physica C357-360, 138 (2001)
\bibitem{27} P.M. Singer, A.W. Hunt, A.F. Cederstr\"{o}m and T. Imai, Phys. Rev. B60, 15345 (1999-II)
\bibitem{28} S. Dzhumanov, Solid State Commun. 115, 155 (2000)
\bibitem{29}A. Damascelli, Z. Hussian and Z.-X. Shen, Rev. Mod. Phys. 75, 473 (2003)
\bibitem{30} L.M. Falicov, IEEE Quantum Electronics. 35, 2358 (1989)
\bibitem{31}H.E. Hussey and J.R. Copper, IRC Research Review 1998.
High Temperature Superconductivity, edited by W.Y. Liang (Cambridge
Univercity Press, Cambridge, 1998) P.52
\bibitem{32} Y. Toyozawa, Technical Report ISSP, Japan. Ser. A, N648 (1974)
\bibitem{33} D. Emin and Holstein, Phys. Rev. Lett. 36, 323 (1976)
\bibitem{34} Y. Toyozawa, Physica B 116, 7 (1983)
\bibitem{35} P.K. Khabibullaev and S. Dzhumanov, Dokl. Akad. Nauk
USSR. 267, 1361 (1982)
\bibitem{36} S. Dzhumanov, preprint, IC/IR/2001/17 (Trieste, Italy)
\bibitem{37} J.T. Markert, Y. Dalichaoch and M.B. Maple, in Physical
Properties of High Temperature Superconductors I, edited by D.M.
Ginsberg (Mir, Moscow, 1990) P.265
\bibitem{38} N.F. Mott and E.A. Devis, Electronic Proceses in Non-Crystaline Materials (Mir, Moskow, 1982)
\bibitem{39} D. Emin, private communication
\bibitem{40} J. Appel, in Polarons, edited by Ya.A. Firsov (Nauka,
Moscow, 1975)
\bibitem{41} B.K. Ridley, Quantum Processes in Semiconductors (Mir, Moscow, 1986)
\bibitem{42} J.P. Falck, A. Levy, M.A. Kastner and R.J. Birgeneau,
Phys. Rev. Lett. 69 1109 (1992)
\bibitem{43}A. Lanzara, P.V. Bogdanov, S.A. Kellar, X.J. Zhou, E.D.
Lu, G. Gu, J.-I. Shimoyama, K. Kishio, Z. Hussain and Z.-X. Shen, J.
Phys. Chem. Solids 62, 21 (2001)
\bibitem{44} Ch. Lushchik, I. Kuusmann, E. Feldbach, F. Savikhin, I. Bitov, J.
Kolk, T. Leib, P. Liblik, A. Maaroos and I. Meriloo, Proc Inst.
Phys. Acad. Sci. Est. SSR. 63, 137 (1987)
\bibitem{45} J.P. Lu and Q. Si, Phys. Rev. B 42, 950 (1990)
\bibitem{46} R.C. Baetzold. Phys. Rev. B 42, 56 (1990)
\bibitem{47} D. Emin, Adv. Phys. 22, 57 (1973)
\bibitem{48} Y. Toyozawa and Y. Shinozuka, J.Phys.Soc.Jpn. 48, 472 (1980)
\bibitem{49} H. B\"{o}ttger and V.V. Bryksin, Hopping Conduction in
Solids (Akademie-Verlag, Berlin, 1985)
\bibitem{50} K. Nasu. Phys. Rev. B35, ¹4, P.1748-1763, 1987.
\bibitem{51} A.V. Puchkov, D.N. Basov and T. Timusk, J. Phys.: Condens. Matter 8,
10049 (1996)
\bibitem{52} F. Walz, J. Phys.: Condens. Matter. 14, R285 (2002)
\bibitem{53} T. Holstein, Ann. Phys. (N.Y.) 8, 325 (1959)
\bibitem{54} A. Ino, C. Kim, M. Nakamura, T. Yoshida, T. Mizokawa,
A. Fujimori, Z.-X. Shen, T. Kakeshita, H. Eisaki and S. Uchida,
Phys. Rev. B65, 094504 (2002)
\bibitem{55} Sh. Sakita, F. Nakamura, T. Suzuki and T. Fujita J. Phys. Soc. Jon. 68, 2755 (1999)
\bibitem{56} Y. Maeno, N. Kakehi,M. Kato, Y. Tanaka and T. Fujita, Physica C185-189, 909 (1991)
\bibitem{57} Y. Maeno, N. Kakehi, M. Kato and T. Fujita, Phys. Rev. B44, 7753
(1991)
\bibitem{58} Verbist, F.M. Peeters and J.T. Devreese, Solid State
Commun. 76, 1005 (1990); Phys. Rev. B43, 2712 (1991)
\bibitem{59} P.J. Baimatov, D.Ch. Khuzhakulov and Kh.T. Sharipov,
Fiz. Tverd. Tela 39, 284 (1997)
\bibitem{60} S. Dzhumanov, P.J.Baymatov, N.P. Baymatova, Sh.T. Inoyatov and O. Ahmedov, eprint,
arXiv:0909.2414
\bibitem{61} S. Dzhumanov, A.A. Baratov and S. Abboudy, Phys. Rev. B54, 13121
(1996-II)
\bibitem{62} S. Dzhumanov, B. Yavidov and N.A. Makhmudov,
Superlattices and Microstructures. 21, 325 (1997)
\bibitem{63} Wu Xiaoguang, F.M. Peeters and J.T.Devreese, Phys. Rev.
31, 3420 (1985)
\bibitem{64} D. Emin, Phys. Rev. B52, 13874 (1995-I)
\bibitem{65} D. Emin, in Polarons and Bipolarons in High-$T_c$
Superconductors and Related Materials, ed. E.K.H. Salje, A.S.
Alexandrov and W.Y. Liang (Cambridge University Press, Cambridge,
1995) P.80
\bibitem{66}S. Dzhumanov, P.J. Baimatov, A.A. Baratov and N.I.
Rahmatov, Physica C235-240, 2339 (1994)
\bibitem{67} S. Dzhumanov and P.K. Khabibullaev Pramana - J. Phys.
45, 385 (1995)
\bibitem{68}  D. Emin, in: Models and Methods of High-$T_c$ Superconductivity,
V. 2, edited by J.K. Srivastava and S.M. Rao (Nova Science
Publisher, Inc., 2003), pp. 343-367
\end{thebibliography}

\end{document}